\documentclass[useAMS,usenatbib,usegraphicx]{mn2e}

\usepackage{setspace}
\usepackage{fixltx2e}

\newcommand{\lya}{{\rm Ly}\alpha}

\newcommand{\hkpc}{h^{-1}{\rm kpc}}
\newcommand{\hmpc}{h^{-1}{\rm Mpc}}
\newcommand{\mpc}{{\rm Mpc}}

\newcommand{\kms}{\;{\rm km}\,{\rm s}^{-1}}

\newcommand{\cmc}{\;{\rm cm}^{-3}}

\newcommand{\msolar}{\;{\rm M}_{\odot}}

\newcommand{\vw}{{v_{\rm wind}}}

\newcommand{\gad}{{\sc Gadget-2}}

\newcommand{\CIV}{\hbox{C\,{\sc iv}}}

\newcommand{\OVI}{\hbox{O\,{\sc vi}}}

\title[Feedback \& Recycled Wind Accretion]{Feedback and Recycled Wind Accretion: Assembling the $z=0$ Galaxy Mass Function} 

\author[B. D. Oppenheimer et al.]{ 
\parbox[t]{\textwidth}{\vspace{-1cm}
Benjamin D. Oppenheimer$^{1,2}$, Romeel Dav\'e$^2$, Du\v{s}an Kere\v{s}$^3$,
Mark Fardal$^4$, Neal Katz$^4$, Juna A. Kollmeier$^5$, David H. Weinberg$^{6,7}$}
\\\\$^1$ Leiden Observatory, Leiden University, PO Box 9513, 2300 RA Leiden, the Netherlands
\\$^2$ Astronomy Department, University of Arizona, Tucson, AZ 85721, USA
\\$^3$ Harvard-Smithsonian Center for Astrophysics, Cambridge, MA 02138, USA
\\$^4$ Astronomy Department, University of Massachusetts, Amherst, MA 01003, USA
\\$^5$ Observatories of the Carnegie Institution of Washington, Pasadena, CA 91101, USA
\\$^6$ Astronomy Department, Ohio State University, Columbus, OH 43210, USA
\\$^7$ Institute for Advanced Study, Princeton, NJ 08450, USA
}

\begin{document}

\pagerange{\pageref{firstpage}--\pageref{lastpage}} \pubyear{2009}

\maketitle

\label{firstpage}

\begin{abstract}
  We analyse cosmological hydrodynamic simulations that include
  theoretically and observationally motivated prescriptions for
  galactic outflows. If these simulated winds accurately represent
  winds in the real Universe, then material previously ejected in
  winds provides the dominant source of gas infall for new star
  formation at redshifts $z<1$.  This recycled wind accretion, or {\it
  wind mode}, provides a third physically distinct accretion channel
  in addition to the ``hot'' and ``cold'' modes emphasised in recent
  theoretical studies.  The recycling time of wind material ($t_{\rm
  rec}$) is shorter in higher-mass systems owing to the interaction
  between outflows and the increasingly higher gas densities in and
  around higher-mass halos.  This {\it differential recycling} plays a
  central role in shaping the present-day galaxy stellar mass function
  (GSMF), because declining $t_{rec}$ leads to increasing wind mode
  galaxy growth in more massive halos.  For the three feedback models
  explored, wind mode dominates above a threshold mass that primarily
  depends on wind velocity; the shape of the GSMF therefore can be
  directly traced back to the feedback prescription used.  If we
  remove all particles that were ever ejected in a wind, then the
  predicted GSMFs are much steeper than observed.  In this case galaxy
  masses are suppressed both by the ejection of gas from galaxies and
  by the hydrodynamic heating of their surroundings, which reduces
  subsequent infall.  With wind recycling included, the simulation
  that incorporates our favoured momentum-driven wind scalings
  reproduces the observed GSMF for stellar masses $10^9 \msolar \leq M
  \leq 5\times 10^{10} \msolar$.  At higher masses, wind recycling
  leads to excessive galaxy masses and star formation rates relative
  to observations.  In these massive systems, some quenching mechanism
  must suppress not only the direct accretion from the primordial IGM
  but the re-accretion of gas ejected from star-forming galaxies.  In
  short, as has long been anticipated, the form of the GSMF is
  governed by outflows; the unexpected twist here for our simulated
  winds is that it is not primarily the ejection of material but how
  the ejected material is re-accreted that governs the GSMF.
 \end{abstract}


\section{Introduction}  

A fundamental challenge in galaxy formation theory is to understand
the mechanisms that shape the observed galaxy stellar mass function
(GSMF).  This requires understanding how galaxies form and grow within
dark matter halos.  Observations and theory indicate that the relation
of galactic stellar components to underlying dark matter halos is not
simple.  In the mass range corresponding to halos of individual
galaxies, the dark matter halo mass function, $\Phi(M_{\rm h})$,
follows roughly $dn/dM_{\rm h}\sim M_{\rm h}^{-2}$
\citep[e.g][]{she99,jen01,spr05b}, turning over only in the group mass
range.  On the other hand, the GSMF, $\Phi(M_{\rm s})$, is most often
fit using a Schechter function, $dn/dM_{\rm s}\sim M_{\rm s}^{\alpha}
{\rm exp}(-M_{\rm s}/M^*)$ \citep{sch76}, where $M_{\rm s}$ is stellar
mass and $z=0$ $M^*$ is roughly the mass of the Milky Way.  The
low-mass slope $\alpha$ is much shallower than $-2$
\citep[e.g.][]{bal08}.  Hence the efficiency of star formation
decreases at masses both lower and higher than $M^*$.  {\it A theory
of galaxy formation must explain the following features of the GSMF: a
{\it preferred} mass for star formation efficiency, a shallow low-mass
end slope, and a rapid drop at large masses.}

We categorise physical mechanisms that imprint features on the GSMF
into two classes: those having to do with how galaxies obtain baryons
(``accretion''), and those having to do with how galaxies lose baryons
(``feedback'').  For accretion, the cooling of the hot virialised
atmospheres, i.e. ``hot mode'' accretion, was historically believed to
be the main channel of gas supply to galaxies.  Models of this process
give a characteristic mass above which gas in a gravitationally
contracting halo cannot cool in a Hubble time ($t_{\rm H}$) to form
stars \citep{ree77, sil77, whi78}, thereby providing a turnover in the
theoretical GSMF, though in the absence of feedback the implied mass
scale is higher and the cutoff shallower than observed \citep{whi91,
  tho95, ben03}.

More recently, ``cold mode'' accretion was argued to be responsible
for most of the gas supply into galaxies across cosmic time
(\citealt{kat03, ker05}, hereafter K05; \citealt{dek06}).  In cold
mode, gas is supplied via cold filaments that rapidly stream gas from
the intergalactic medium (IGM) into star-forming regions, without ever
encountering a virial shock~\citep{bin77, bir03}.  This is most
effective in lower-mass halos, while in halos $\ga 10^{12} \msolar$
shocked gas dominates the halo gas, although dense filaments at high
redshifts ($z\ga 2$) still allow cold mode accretion to proceed even
above this threshold mass (K05; Dekel \& Birnboim 2006; Kere\v{s}
2007; Ocvirck et al. 2008; Kere\v{s} et al. 2009a, hereafter K09a;
Brooks et al. 2009; Dekel et al. 2009).  At lower redshifts, the
supply of cold gas declines in massive halos, and the cooling of the
hot atmosphere becomes relatively more important (K09a).  This late
hot mode is relatively inefficient and can be naturally prevented in
the large, constant density cores of cluster halos.

Yet accretion alone cannot be the entire story, as it is well known
that without strong feedback, models fail to produce basic
observations such as the luminosity function and the GSMF
\citep{kau93, col94, mur02, bow06}.  \citet[][hereafter K09b]{ker09b}
explored a cosmological SPH simulation without any form of explicit
feedback and derived a ``correction factor'' as a function of stellar
mass to correct the simulated GSMF to match the observed one.  The
fact that their correction factor is always greater than one
exemplifies the ``over-cooling'' problem in galaxy formation, where
the efficiency of forming stars, based on simple theoretical
expectations, is greater than that actually observed
\citep[e.g.][]{whi91,bal01}.  {\it Distinct feedback processes may
  operate at different mass scales}, given that star formation needs
to be suppressed by $\sim 10$ at $M_{\rm s}<10^{10} \msolar$, falling
to only $\sim 2$ at $10^{11} \msolar$, and then increasing again to
$\sim 5$ by $10^{12} \msolar$ (see $f_{\rm corr}$ plotted in K09b,
Figure 1).

K09b differentiated ``preventive'' feedback, where a process prevents
gas from entering a galaxy's ISM in the first place, and ``ejective''
feedback, where gas is accreted and then thrown back out. Preventive
feedback is most applicable for hot mode, since halos with dilute hot
atmospheres can be more effectively kept hot by feedback processes
such as jets from active galactic nuclei (AGN) and magnetic conduction
\citep[e.g.][K05]{bin04}.  An example of preventive feedback affecting
hot mode is ``radio''-mode feedback~\citep{bes05,cro06}.  However,
K09b showed that preventing hot mode accretion alone is insufficient
to correct the GSMF, since in large systems most of the stars are
still formed via cold mode when the galaxy was smaller (K09a), meaning
that {\it cold mode must be suppressed}.  Preventing cold mode
accretion is much more challenging, since this gas is streaming in via
dense filaments that are not easily disrupted.  Hence, even for
galaxies that are today dominated by hot mode accretion, ejective
feedback appears necessary to understand their present-day stellar
masses.

In short, both the high and low ends of the observed GSMF appear to
{\it require an ejective feedback mechanism} to curtail cold
mode-driven star formation.  A leading candidate for such a mechanism
is galactic winds \citep[e.g.][]{dek86}.  Winds ejecting galactic ISM
have the added benefit of chemically enriching the IGM, where metal
absorption lines are observed to be common
\citep[e.g.][]{son96,son01,sch03,sim04,ade05}.  Star formation
suppression, IGM enrichment, plus the ubiquity of escaping outflows
observed in star forming galaxy spectra \citep[e.g.,][]{pet01, sha03,
  mar05a, wei09} have motivated the inclusion of star formation based
feedback prescriptions in cosmological simulations \citep{spr03b,
  kob04, cen06a, opp06, som08, wie09b}.  \citet{opp06} explored a
range of feedback prescriptions, finding that the scalings predicted
by a momentum-driven wind model \citep{mur05} were most promising for
matching the observations of high-$z$ IGM enrichment.  In this
scenario, wind velocity ($\vw$) scales with galaxy velocity dispersion
($\sigma$), leading to the gradual enrichment of the IGM matching the
observed frequency of $z\sim 6$ $\CIV$ lines \citep{opp09b}, the
evolution of $\CIV$ from $z=5\rightarrow 2$ \citep{opp06}, plus the
kinematics of damped $\lya$ absorbers \citep[DLAs,][in
preparation]{hon09}.  The mass loading factor ($\eta$, where
$\dot{M}_{\rm wind}=\eta \dot{M}_{\rm SF}$) scales as $\sigma^{-1}$,
meaning that lower mass galaxies are the most prolific enrichers by
mass.  This relation is central to reproducing the $z\sim 6$ galaxy
luminosity function \citep{dav06}, explaining the galaxy
mass-metallicity relation \citep{fin08a}, and providing a solution to
the missing halo baryon problem \citep{dav09}.  The prescription even
appears important for systems in the local Universe, as these wind
simulations can account for the $z=0$ missing metals \citep{dav07},
the observed $\OVI$ forest at $z<0.5$ \citep{opp09a}, and intragroup
enrichment and entropy levels at $z=0$ \citep{dav08b}.  While our
simulations are only sensitive to the predicted wind scalings, and do
not try to model the detailed physics of wind propagation, there is
independent evidence that other fundamental properties of galaxies are
best understood by invoking momentum-driven winds~\citep{hop10}.

One consequence of having highly mass-loaded, moderate-velocity
outflows emanating from lower mass galaxies is that wind material
remains in the relative proximity of their parent galaxies.
\citet[][hereafter OD08]{opp08} found that even when outflows launch
in excess of a halo's escape velocity, wind material regularly
``recycles'' back into a galaxy on a median timescale of 1-2 Gyrs.
This occurs because winds are more slowed hydrodynamically by
interactions with halo gas and gas outside halos rather than
gravitationally.  Winds typically reach $\sim 100$ physical kpc at all
epochs, which means that outflows from $\ge M^*$ galaxies do not
escape from their halo after $z\sim 1$.  OD08 introduced the term {\it
halo fountain} to describe such winds that remain trapped in their
parent halos~\citep[see also][]{ber07}.  K09b coined {\it
intergalactic fountain} to describe the transfer of feedback gas from
low-mass galaxies at early times to higher mass galaxies at later
times via a journey through the IGM.

A major implication of our wind models is that the amount of material
ejected from galaxies is, globally over cosmic time, far larger than
the amount of material forming into stars.  Hence these outflow
fountains recycling back into galaxies can in principle represent a
significant reservoir for stellar mass formation over a Hubble
timescale.  This paper is devoted to examining such accretion, which
we call ``recycled wind accretion'' (or just ``wind mode''), in
cosmological hydrodynamic simulations.  In addition to hot and cold
accretion, wind accretion effectively provides a third avenue for how
galaxies get their gas.  In this work we examine issues such as: How
much of the cosmic stellar mass growth can this accretion mode create?
Can this mode explain the preferred mass of galactic star formation at
$M^*$ (i.e. the knee in the Schechter function), as well as the
faint-end slope of the GSMF?  What happens to the GSMF if winds never
recycle?  Does wind mode need to be ``quenched'' at any particular
mass, and if so what physics might produce this quenching?

In \S2 we introduce our latest cosmological hydrodynamic simulations
with several different prescriptions of superwind feedback.  We begin
to answer some the above questions in \S3 when we consider stellar
mass growth divided into these component modes.  We emphasise the
concept of differential wind recycling as a function of galaxy mass
and link it to the mass dependence of wind mode accretion.  The
effects of wind recycling on the GSMF are analysed in
\S\ref{sec:gsmf}.  Our purpose is not to precisely fit the $z=0$ GSMF,
but to introduce the concept of wind mode as a source of galaxy
growth.  We do argue that plausible wind models can come close to
matching the observed GSMF, but only for galaxies below $M^*$.  In \S4
we discuss the balance of ejective and preventive feedback with
recycled wind accretion.  Ignoring wind mode, we examine how feedback
suppresses star formation in \S\ref{sec:suppress}, finding that no
feedback implementation we explore can adequately explain the observed
GSMF without wind mode.  We then emphasise the effect of wind mode on
the slope of the GSMF in \S\ref{sec:faintend}.  \S\ref{sec:numeff}
examines whether wind mode is partly a numerical effect, especially
above $M^*$ where star formation-driven outflows are completely
ineffective at quenching star formation.  Finally, \S5 summaries our
findings.

\section{Simulations} \label{sec:sims}

We employ the \gad~N-body + Smoothed Particle Hydrodynamic (SPH) code
\citep{spr05} to evolve a series of cosmological simulations to $z=0$.
We adopt a $\Lambda$CDM cosmology using cosmological parameters
based on the 5-year WMAP results~\citep{hin09}: $\Omega_{\rm M}=0.28$,
$\Omega_{\rm \Lambda}=0.72$, $h\equiv H_0/(100 \kms \mpc^{-1})=0.7$, a
primordial power spectrum index $n=0.96$, an amplitude of the mass
fluctuations scaled to $\sigma_8=0.82$, and $\Omega_{\rm b}=0.046$.
We refer to these simulations as the r-series, where our general
naming convention for a simulation is r[{\it boxsize}]n[{\it
  particles/side}][{\it wind model}].  Our primary simulations use a
boxsize of 48 $\hmpc$ with 384 dark and SPH particles/side resulting
in a dark matter particle mass of $1.8\times10^8 \msolar$ and an SPH
particle mass of $3.6\times10^7 \msolar$.  Gas particles on average
spawn two star particles resulting in star particle masses averaging
$1.8\times10^7 \msolar$.  We define the wind model suffixes at the end
of this section.

An overview of the \gad~code can be found in \S2 of K09a.  Additions
to the public version of the code include cooling processes using the
primordial abundances as described in \citet{kat96} and metal-line
cooling as described in \citet{opp06}.  Star formation is modelled
using a subgrid recipe introduced in \citet{spr03a} where a gas
particle above a critical density is modelled as a fraction of cold
clouds embedded in a warm ionised medium as in \citet{mck77}.  The
star formation follows a Schmidt law \citep{sch59} with the star
formation timescale scaled to match the $z=0$ Kennicutt law
\citep{ken98}.  The density threshold for star formation is $n_{\rm
  H}=0.13 \cmc$.

We use a \citet{cha03} initial mass function (IMF) throughout, which
has a turnover at masses $<1 \msolar$ relative to the \citet{sal55}
IMF.  The fraction of mass in the IMF going into massive stars,
defined here as $\ge 10 \msolar$ and assumed to result in Type II
supernovae (SNe), is $f_{\rm SN}=0.18$ in the Chabrier IMF.  As our
$f_{\rm SN}$ is larger than for the Salpeter IMF used in
\citet[][$f_{\rm SN}=0.10$]{spr03a}, the amount of SN feedback energy
coupled to the ISM is greater per unit star formation as is the energy input
into galactic outflows in our standard constant-wind model (see
below).

Our simulations directly account for metal enrichment from sources
including Type II SNe, Type Ia SNe, and AGB stars.  Gas particles
eligible for star formation undergo self-enrichment from Type II
supernovae (SNe) using the instantaneous recycling approximation,
where mass, energy, and metallicity are assumed to immediately return
to the ISM.  Type II SN metal enrichment uses the
metallicity-dependent yields calculated from the \citet{chi04} SNe
models.  Our prescriptions for feedback, described below, also assume
the instantaneous input of energy from O and B stars (i.e. stars that
are part of $f_{\rm SN}$).  We input the Type Ia SNe rates of
\citet{sca05}, where an instantaneous component is tied to the SFR,
and a delayed component is proportional to the stellar mass as
described in OD08.  Each Type Ia SN adds $10^{51}$ ergs of thermal
energy and the calculated metal yields of \citet{thi86} to surrounding
gas, while the mass returned is comparatively negligible.  AGB stellar
enrichment occurs on delayed timescales from 30 Myrs to 14 Gyrs, using
a star particle's age and metallicity to calculate mass loss rates and
metallicity yields as described in OD08.  AGB mass and metal loss is
returned to the three nearest surrounding gas particles.  OD08 showed
that the largest effect of AGB stars is to replenish the ISM, because
a star particle can lose nearly 40\% of its mass over $t_{\rm H}$,
about double the $f_{\rm SN}=18\%$ that is recycled instantaneously
via Type II SNe.

Feedback is directly tied to the SFR, using the relation $\dot M_{\rm
wind}= \eta \dot M_{\rm SF}$, where $\eta$ is the ``mass loading
factor.''  The probabilities for an SPH particle with a non-zero SFR
to turn into a star particle and to join a wind are calculated in each
timestep; the ratio of the two probabilities is therefore $\eta$.
Actual conversion into a star or wind particle is determined
stochastically using these probabilities and a random number
generator.  We explore simulations with no winds and three different
forms of implemented feedback.

{\bf No Winds (nw):} No feedback energy is imparted kinetically to SPH
particles.  However, as with all simulations, energy is imparted
thermally to ISM SPH particles using the \citet{spr03a} subgrid
two-phase recipe where all SN energy is instantaneously coupled to the
ISM.  The r48n384nw simulation can be compared to the \gad~no wind
simulation explored in K09a and K09b, although there are some notable
differences.  First, the cosmology is different; they use $\Omega_{\rm
  M}=0.26$, $\Omega_{\rm \Lambda}=0.74$, $h=0.71$, $n=1.0$,
$\sigma_8=0.75$, and $\Omega_{\rm b}=0.044$.  Second, the r-series uses a
Chabrier IMF with delayed mass loss from AGB stars and feedback from
Type Ia SNe, while the K09a simulation uses a Salpeter IMF and only
Type II SNe feedback.  The recycling of gas from AGB stars causes more
late-time star formation in the $r$-series. A third difference is the
addition of metal-line cooling for non-star forming SPH particles.
This actually has little effect in the no wind case, because a vast
majority of metals ($\ge 96\%$) remain within galaxies in either the
ISM or star particles. We will find throughout \S3 that the r48n384nw
simulation reproduces the main mass growth properties (i.e.  cold \&
hot mode) of the 50 $\hmpc$, $2\times 288^3$ particle simulation of
K09a/b, although with some minor differences.

{\bf Constant Winds (cw):} Feedback energy is added kinetically to gas
particles at a constant rate relative to the SFR, with $\eta=2$ and a
constant velocity $\vw=680 \kms$.  This is intended to mimic the
constant-wind model of \citet{spr03a}, however with some notable
differences.  Here $\eta$ is defined relative to the SFR of the entire
Chabrier IMF, and not just the long-lived stars in a Salpeter IMF as
it is defined by \citet{spr03a}.  Second, $\vw=680 \kms$ is used
rather than $484 \kms$, because there is more SN energy per mass
formed in a Chabrier IMF.  The kinetic energy imparted to wind
particles per unit star formation is $9.25\times 10^{48} {\rm
erg}/\msolar$, which is 95\% of the SN energy in this IMF assuming all
stars $\ge 10 \msolar$ add $10^{51}$ ergs/SN.  Wind particles are
hydrodynamically decoupled until either $2.9\times10^7$ yrs has passed
or, more often, the SPH particle has reached 10\% of the star
formation critical density.  \citet{spr03b} demonstrated that this
decoupling achieved resolution convergence. It is supposed to simulate
the formation of hot chimneys extending out of a disk galaxy providing
a low resistance avenue for SN feedback to escape into the IGM.

{\bf Slow Winds (sw):} We simulate an alternative constant-wind model with
the only difference being the outflow velocities are half as high as
the cw winds (i.e. $\vw=340 \kms$ and $\eta=2$).  Therefore, only a
quarter of the SN energy couples kinetically into these outflows.

{\bf Momentum-driven Winds (vzw):} The momentum-driven wind model uses
the scalings of \citet{mur05} based on the galaxy velocity dispersion
($\sigma$), and matches the observed trends of $\vw$ by
\citet{mar05a}.  Its implementation is explained in \citet{opp06} and
updated in OD08 with the addition of an in-run group finder to
calculate $\sigma$ using galaxy mass following \citet{mo98}.  One
modification from OD08 is the in-run group finder now uses dark matter
in addition to baryons to calculate masses as this produces better
agreement across time with our post-run group finder (discussed
below).  The wind parameters vary across galaxies using the following
relations:
\begin{eqnarray}
  \vw &=& 3\sigma \sqrt{f_{\rm L}-1}, \label{eqn:windspeed} \\
  \eta &=& {\sigma_0\over \sigma} \label{eqn:massload},
\end{eqnarray} 
where $f_{\rm L}$ is the luminosity factor in units of the galactic
Eddington luminosity (i.e. the critical luminosity necessary to expel
gas from the galaxy potential), and $\sigma_0$ is the normalisation of
the mass loading factor.  Here we include a metallicity dependence
additionally for $f_{\rm L}$ owing to more UV photons being output by lower
metallicity stellar populations
\begin{equation}
  f_{\rm L} = f_{\rm{L},\odot} \times 10^{-0.0029*(\log{Z}+9)^{2.5} + 0.417694},
  \label{eqn:zfact}
\end{equation}
randomly select $f_{\rm{L}, \odot}=[1.05,2]$ for each SPH particle,
following observations by \citet{rup05}, and add an extra kick of size
$2\sigma$ (the local escape velocity) to simulate continuous pumping
of winds until it is far from the galaxy disk (as argued by
\citet{mur05}).  The only free parameter is $\sigma_0=150 \kms$, which
is the same efficiency factor of the mass loading used in OD08.
Similar to the cw model, hydrodynamic forces are decoupled for a time
of $1.95\times10^{10} \kms/ \vw$ yr or until 10\% of the star
formation critical density is reached; the latter is the case for the
majority of wind particles.

The in-run group finder used to derive wind properties produces
robustly consistent masses used to calculate $\sigma$ compared to the
post-run halo finder, although there are differences.  The velocity
dispersion of a halo, $\sigma_{\rm h}$, scales as $M_{\rm h}^{1/3}$,
where $M_{\rm h}$ is derived by post-run.  $\sigma_{\rm h}$ is
different than $\sigma$ calculated by the in-run group finder, which
uses a higher threshold density to identify individual galaxies in
groups and clusters rather than the dispersion of the entire
group/cluster halo.  Therefore, we find simulated winds have median
values following $\eta\propto \sigma_{\rm h}^{-0.95}$ between $M_{\rm
h}=10^{10.7}-10^{13.3} \msolar$.  The $\vw$ medians scale as
$\sigma_{\rm h}^{0.83}$, deviating from a linear relation owing
primarily to lower metallicities from low-mass galaxies resulting in
an extra boost (Equation \ref{eqn:zfact}); $\eta$ does not depend on
this equation.  Because $\vw$ does not linearly scale with
$\sigma_{\rm h}$, winds launched from galaxies residing in groups and
clusters are often launched below the halo escape velocity ($v_{\rm
esc}$); we explore this further in \S\ref{sec:diffrec}.  At $z=1$, a
$10^{12} \msolar$ halo identified by the post-run halo finder produces
winds with median $\vw=490 \kms$ and $\eta=1.7$, equivalent to 40\% of
the total SN energy.

Our four simulations are listed in Table \ref{tab:sims}.  They were
run at the University of Arizona's SGI cluster, ICE, the National
Center for Supercomputing Applications' Intel cluster, Abe, and the
University of Massachusetts' Opteron cluster, Eagle.  Each simulation
took between 100,000 and 150,000 CPU hours, although nearly 1 million
CPU hours were used when including smaller box sizes and various test
simulations.

\begin{table}
\caption{Simulations$^a$}
\begin{tabular}{lccc}
\hline
Name &
$\vw$ ($\kms$) & 
$\eta^{b}$ &
$E_{\rm wind}/E_{\rm SN}^{c}$
\\
\hline
\multicolumn {4}{c}{} \\
r48n384nw   & -- & -- & --   \\
r48n384cw   & 680 & 2 & 0.95 \\
r48n384sw   & 340 & 2 & 0.24 \\
r48n384vzw  & $\propto\sigma$ & $\propto\sigma^{-1}$ & 0.72 \\
\hline
\end{tabular}
\\ 
$^a$ All simulations have $384^3$ SPH and dark matter particles in a
48 comoving $\hmpc$ box, corresponding to $1.8\times10^8$ and
$3.6\times10^7 \msolar$ per SPH and dark matter particle,
respectively.  They were all evolved to $z=0$.  The equivalent Plummer
gravitational softening length is 2.5 comoving $\hkpc$.\\
$^b$ $\eta$ is defined as $\dot{M}_{\rm wind} \propto \eta\times M_{\rm SF}$.\\
$^c$ Feedback energy divided by SN energy.  Averaged over resolved galaxies for the r48n384vzw simulation.  
\\

\label{tab:sims}
\end{table}

We use the program
SKID\footnote{http://www-hpcc.astro.washington.edu/tools/skid.html} to
identify bound groups of cold baryons and stars (see K05 and K09a for
more details).  We have modified SKID with the requirement that bound
gas particles must have non-zero SFRs; therefore we only identify gas
in the galactic ISM and not from the extended halo.  Our galaxy
stellar mass limit is defined as $\ge 64$ star particles, $M_{\rm
  s}\sim 10^{9.05} \msolar$, because \citet{fin06} showed stellar
properties of galaxies are well-converged above this limit.  Halos are
identified using a Spherical Overdensity algorithm with the same
parameters as in K09a.

\section{Results}

\subsection{Modes of Star Formation} \label{sec:modes}

We begin our discussion of stellar mass growth by dividing gas accretion
and star formation into its component modes.  Just as in K05 \& K09a/b, we
divide hot and cold mode based on the maximum temperature, $T_{\rm max}$,
an SPH particle achieves on its journey into a galaxy.  The temperature
split is the same -- $T=2.5\times 10^5$ K, which K05 justified as a good
empirical division of the two modes.  We define a third mode, {\it wind
mode}, as any particle previously ejected in a wind that is re-accreted
by a galaxy and forms stars (i.e. recycled).  There is no temperature
constraint on wind mode.

We have modified our version of \gad~to track these three modes 
during a simulation using the following rules.  An SPH particle's
$T_{\rm max}$ is tracked every timestep if it has not yet been
launched in a wind.  If the particle's current temperature is greater
than $T_{\rm max}$, then $T_{\rm max}$ is set to the current
temperature.  $T_{\rm max}$ cannot be modified if a particle has a
non-zero SFR, where the temperature is an average of the two-phase ISM
model of \citet{spr03a}.  Particle wind launches are also tracked, and
$T_{\rm max}$ is not allowed to be adjusted after such an event.  Any
star formation from a previously wind-launched particle is considered
wind mode.  For the purposes of this paper, we do not consider a
particle's $T_{\rm max}$ nor subsequent temperature history once it is
launched in a wind.  We also cannot properly track gas transferred via
AGB feedback from star particles to surrounding gas particles,
although this is a second-order effect that is not expected to favour
any one mode; the transferred material will inherit the gas particle's
mode.

\begin{figure}
\includegraphics[scale=0.8]{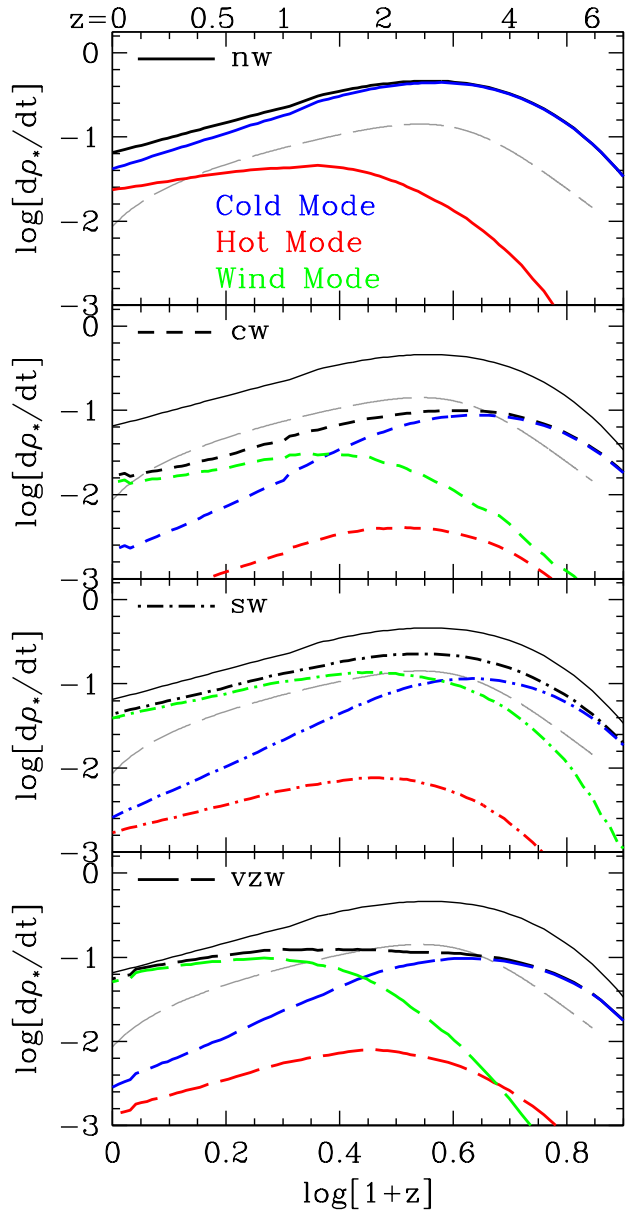}
\caption[]{The global SFR density ($\msolar$ yr$^{-1}$ Mpc$^{-3}$) in
  comoving units as indicated by the black lines, and sub-divided by
  accretion mode: cold (blue), hot (red), and wind (green).  The
  integrated global no-wind SFR density is repeated with thin lines in
  the bottom three panels for comparison with each wind model.  The
  grey dashed line indicates the Hopkins \& Beacom. (2006) compilation
  calibrated to a Chabrier IMF.  All galaxies (centrals and
  satellites) are included. }
\label{fig:madau}
\end{figure}

We consider the global star formation divided into these three modes
for our four models in Figure \ref{fig:madau}.  Our baseline for
comparison is the no-wind (nw) simulation in the top panel.  The black
line is the global SFR density as a function of redshift, and the blue
and red curves correspond to the hot and cold modes of star formation
respectively; we use this colour scheme for these modes throughout
this paper.  The nw model is the focus of K09a and K09b, who use a
lower resolution \gad~simulation without winds as mentioned in
\S\ref{sec:sims}.  Our nw simulation shows the same qualitative
characteristics identified by K09a including: (i) cold mode accretion
dominates at all epochs, and (ii) the hot mode accretion fraction
increases at late times; we refer the reader to their \S3 for an
in-depth discussion.  However, note that Figure~\ref{fig:madau} plots
star formation and not accretion (as was done in Figure 2 of K09a),
although the two are tightly coupled (K05).  Throughout this paper we
consider only SFRs and not accretion rates, as our focus is on stellar
mass growth; we leave a direct investigation of accretion rates for
future work.

The next three panels of Figure \ref{fig:madau} show the global SFR
densities divided by mode for our three feedback models; the thin
black line is the nw global SFR shown for comparison.  In addition to
the hot and cold modes indicated by the red and blue lines, the green
lines in each of these panels indicate the contribution from the
recycled wind mode.  In every case, the wind mode dominates the global
SFR over most of the history of the Universe.  The wind mode
contributes over 50\% of the star formation after $z=1.3$, $2.7$, and
$1.7$ for the cw, sw, and vzw models, respectively.  The wind mode
grows relative to the other modes at late times, making up 84\%, 90\%,
and 92\% of the $z=0$ star formation, respectively.  The resulting
integrated stellar mass fractions in $z=0$ galaxies are 51\%, 73\%,
and 71\%, respectively.

This leads us to our first main conclusion: wind material not only
commonly recycles back into galaxies, it in fact {\it dominates}
late-time stellar mass growth.  Ejected material cannot be assumed to
remain in the IGM forever.  The wind mode and the total SFRs decline
at late times, albeit at a slower rate than the cold and hot modes.
SFRs are lower for every feedback simulation relative to the nw
simulation, but the reduction is fairly minor for sw winds and vzw
winds at later times.  Outflows efficiently suppress both the cold and
hot modes in all cases, but the wind mode re-supplies some of the
ejected gas back into galaxies.  Compared with the observed SFR
density \citep{hop06} (grey dashed lines), these two models
over-predict star formation after $z>1$, which we will argue in
\S\ref{sec:numeff} is primarily a result of excessive star formation
in $>M^*$ galaxies.

\subsection{Differential Wind Recycling} \label{sec:diffrec}

Wind material recycles back into galaxies in a manner that depends on
galaxy mass (OD08).  This trend, which we refer to as {\it
differential recycling}, is key to understanding how recycling shapes
the galaxy stellar mass function in these simulations.  Wind recycling
is defined as an event where a previously launched wind particle is
either launched in a subsequent wind or converted completely into a
star particle.  The recycling time $t_{\rm rec}$ is defined as the
time between the initial launch and re-launch/star particle
conversion, which includes the time spent going out of the galaxy, the
time spent coming back into the galaxy, and that spent within the new
galaxy's ISM.  OD08 showed that individual wind particles average 2-3
wind launches (i.e. 1-2 recyclings) in their vzw simulations with a
median $t_{\rm rec}$ of $\la 1$ Gyr, showing a weak dependence with
redshift.

OD08 also showed recycling in the vzw model was highly dependent on a
galaxy's baryonic mass, with $t_{\rm rec}$ scaling as $\sim M_{\rm
b}^{-1/2}$ for vzw outflows.  At face value this is surprising, since
in this model winds are launched at speeds nearly proportional to
$v_{\rm esc}$ of the galaxy.  Hence if gravity were the dominant
criterion for how far a wind travels and $v_{\rm wind} < v_{\rm esc}$,
one should expect almost no trend with mass.  By comparing the trends
with analytic expectations, OD08 argued that hydrodynamical slowing
dominates the wind deceleration, not gravitational forces.  The
deceleration is greater for more massive galaxies because they reside
in denser gas environments, leading to quicker recycling.  OD08
demonstrated that the gas outside halos is primarily responsible for
hydrodynamic slowing in lower mass halos, while winds rarely escape
the gas-rich halos of $>M^*$ galaxies at $z<1$ (see their Figures 13
and 14).

\begin{figure*}
\includegraphics[scale=0.8]{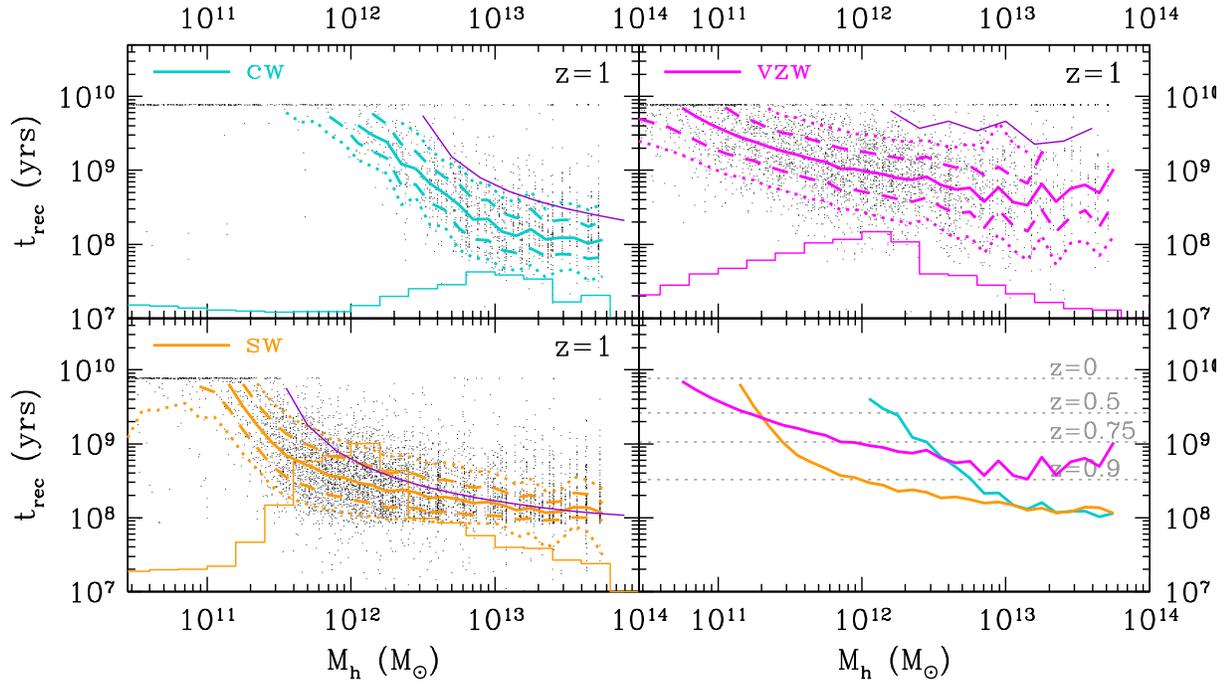}
\caption[]{ Recycling times -- the time between when a wind particle
  is launched and when it is converted to a star or launched for a
  second time -- for all wind particles ejected between $z=1-0.95$ in
  the three wind simulations, as a function of the host galaxy's
  $M_{\rm h}$ at launch time.  Black dots show individual wind
  particles, and coloured solid lines show the median $t_{\rm rec}$;
  if this line exists for a given $M_{\rm h}$ then at least 50\% of
  wind particles recycle.  Similarly, 25 and 75\% cuts on $t_{\rm
  rec}$ are indicated by dashed lines, while 10 and 90\% cuts are
  indicated by dotted lines; if the top dotted line exists, $>90\%$ of
  winds recycle.
  Purple lines show the expected $t_{\rm rec}$ from gravitational
  considerations alone.  Histograms show the relative number of wind
  particles per $M_{\rm h}$ emanating from central galaxies.  Wind
  events that escape until $z=0$ contribute to the thin line of points
  at $\sim 8$ Gyrs, the time between $z=1\rightarrow 0$.  The bottom
  right panel shows the diversity in the differential (i.e.,
  mass-dependent) recycling curves (medians). Dotted grey lines,
  applicable for all models, indicate the redshift when median wind
  particles, launched at $\langle z \rangle=0.975$, recycle. }

\label{fig:diffrec}
\end{figure*}

Figure \ref{fig:diffrec} plots $t_{\rm rec}$ as a function of the halo
mass for central galaxies at the time they launch their winds.  We
focus on wind particles launched between $z=1\rightarrow 0.95$, for
several reasons: (i) most wind mode accretion occurs below $z=1$, (ii)
there is ample time afterward ($\sim 8$ Gyrs) to follow the outflow
material, and (iii) there is a larger dynamic range in $M_{\rm h}$ to
explore dependencies than at higher $z$.  The curves at other
redshifts do not differ dramatically from $z=1$ for the same mass
galaxy (see OD08).  The black points correspond to individual wind
particles; the vertical striping at higher halo masses results from
the finite number of such halos. The un-recycled winds appear as the
thin line of points at $\sim 8$ Gyrs.  The lines show the times below
which 10, 25, 50, 75, and 90\% of launched particles recycle.  If a
line exists at a given $M_{\rm h}$, then at least its corresponding
fraction of particles recycle by $z=0$ (e.g. if top dotted line
exists, $\ge 90\%$ of winds recycle for that $M_{\rm h}$).  Finally,
the histograms along the bottom show the relative number of integrated
wind launches from all halos within a $M_{\rm h}$ bin.  The statistics
here only include central galaxies, which contain 87\% of the total
$z=1$ star formation in each of the three models.  

The purple curves show the analytical $t_{\rm rec}$ expected purely
from gravity (i.e. freely falling trajectories) for each model.  These
are calculated by integrating the time-varying gravitational
acceleration along a particle's trajectory using the $\vw$ as the
launch velocity and assuming \citet{nav97} halo potentials with
concentrations from the \citet{duf08} relation at $z=1$ and a launch
radius of 0.0125 $r_{200}$.  $t_{rec}$ is twice the time needed to
reach the turnaround radius, and is plotted if a particle is
calculated to recycle before $z=0$; no evolution for the halo is
assumed.  The median $t_{\rm rec}$ from the cw and sw models are
shorter by $\times 2-3$ compared to the analytical case owing to
hydrodynamic forces when $t_{\rm rec}\ga 1$ Gyr, confirming the
argument of OD08 that hydrodynamic slowing plays a critical role in
wind recycling.  While winds in the vzw model are launched in excess
of $v_{\rm esc}$ at high-$z$, this is not necessarily the case for
galaxies $\ga M^*$ at $z\la 1$.  Winds launched with $\vw \propto
\sigma$, where $\sigma$ is calculated for an individual galaxy, cannot
escape the group and cluster potentials in which such massive galaxies
typically reside.  We use the median $\vw$ tracked in the simulation
for a given $z=1$ $M_{\rm h}$, therefore there exists some slight
scatter from the expected $\vw-M_{\rm h}$ relation.  This scatter
becomes amplified when performing the calculation of gravitational
$t_{\rm rec}$, leading to the non-monotonic trend of the purple line
in the upper right panel.  Gravitational forces play a significant
role in slowing vzw winds, yet hydrodynamic slowing still reduces
$t_{\rm rec}$ by $\times 3-5$.

The shape of the $t_{\rm rec}$-$M_{\rm h}$ relation for all the wind
models we examined can generally be described by a power law decay of
$t_{\rm rec}$ with mass followed by a gradual flattening.  The
differences between models can be seen explicitly in the bottom-right
panel of Figure \ref{fig:diffrec}, where the median $t_{\rm rec}$
curves are compared.  The cw model exhibits a steeper decline relative
to the vzw model, which is simple to understand.  The cw model has no
dependence of outflow velocity on mass, leading to a greater
gravitational deceleration at higher masses as the wind velocity falls
below the typical $v_{\rm esc}$ of the halo.  Another strong
prediction of the cw model is, as expected, nearly no gas recycling
occurs in halos with masses below $M_{\rm h} \sim 3\times 10^{11}
\msolar$.  This is in sharp contrast to the vzw model where even
low-mass halos experience significant rapid gas recycling.  The sw
model shows similar, but appropriately shifted, behaviour as the cw
model, where the $t_{\rm rec}-M_{\rm h}$ relation has moved to lower
mass corresponding to lower $v_{\rm esc}$.  Quantitatively, $t_{\rm
  rec}\sim M_{\rm h}^{-1.5}$ in cw versus $\sim M_{\rm h}^{-0.6}$ in
vzw over the range where $t_{\rm rec}$ reduces from 5 to 1 Gyr.  Above
$M_{\rm h}\sim 10^{12} \msolar$, the median $t_{\rm rec}$ in the vzw
model flattens and never decreases much below 1 Gyr.  This contrasts
to OD08 where $t_{\rm rec}$ approaches 100 Myrs for the most massive
galaxies. This difference owes to the fact that we impose no speed
limit in the simulations presented here whereas OD08 limits the
$E_{\rm wind}$ to be no more than $2 E_{\rm SN}$, reducing the $\vw$
from massive galaxies.

Finally, the histograms of wind particle launches show significant
differences between the models.  These histograms are tied to the
integrated SFRs per halo mass bin in central galaxies using the
conversion factor $\eta^{-1}$.  For the constant-wind models, this
conversion factor is flat by definition, allowing the histogram to be
a proxy indicator of the integrated star formation in central
galaxies.  The up-tick in the histogram at higher masses corresponds
to the steep reduction in $t_{\rm rec}$ and is the signature of wind
recycling significantly increasing star formation in higher mass halos
at $z=1$.  This up-tick occurs at lower $M_{\rm h}$ in the sw model
relative to the cw model, which we will show is related to the $M_{\rm
  h}$ where the wind mode dominates in the next subsection.  Also,
the integrals of each histogram are comparable between panels, and are
proxies for the global $z=1$ SFRs in Figure \ref{fig:madau} (not
including the minor contribution from satellites).  The higher
histogram of the sw model relative to the cw model results from the
greater $z=1$ wind mode SFRs.  The vzw model cannot be so easily
interpreted since $\eta$ declines with halo mass.

To summarise, the wind velocity primarily sets the recycling time at a
given halo mass, while the mass loading factor multiplied by the SFR
determines the integrated amount of winds emanating from a given halo
mass range (i.e. the histograms in Figure \ref{fig:diffrec}).  The
wind launch histograms increase sharply where $t_{\rm rec}$ becomes
small relative to $t_{\rm H}$, which links differential recycling and
the re-accretion of gas for star formation.  

\subsection{Star Formation and Stellar Masses} \label{sec:SFRs}

\begin{figure*}
\includegraphics[scale=0.8]{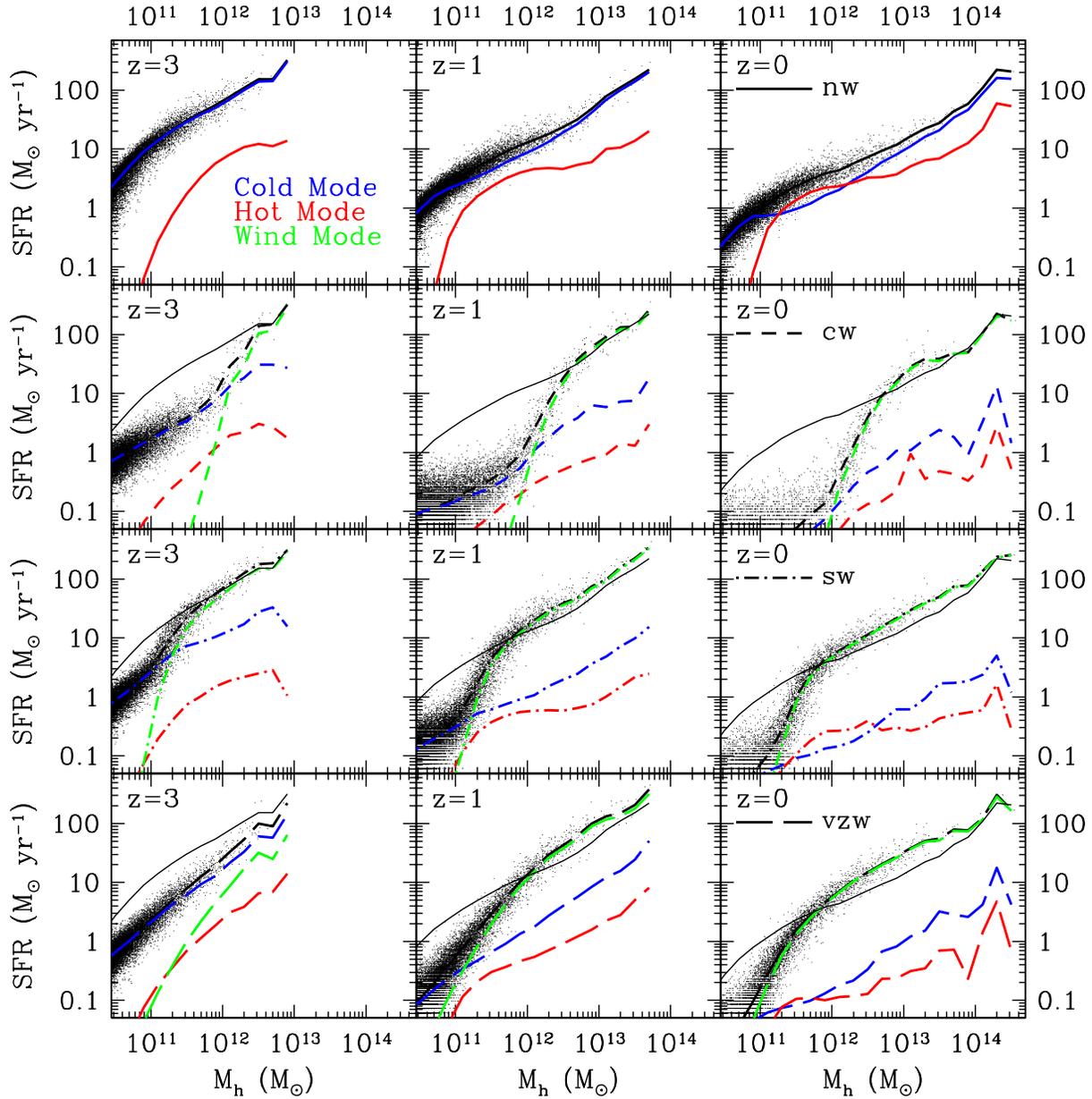}
\caption[]{Instantaneous SFRs of central galaxies as a function of
  $M_{\rm h}$ for three redshifts ($z=3,1,0$) in our four
  simulations.  Black points show total SFRs of individual galaxies,
  and the black line shows the mean.  Blue, red, and green lines show
  the mean SFR for cold, hot, and wind modes respectively.  The thin
  black lines indicating the total nw SFR are repeated in the feedback
  model panels for comparison. 
  }
\label{fig:sfrmode}
\end{figure*}

Having examined how winds recycle, we now consider this gas resupply
as a source of star formation in Figure~\ref{fig:sfrmode}, where the SFR
modes are decomposed as a function of $M_{\rm h}$ in our four
models at three redshifts ($z=$3, 1, and 0).  The data points represent
individual galaxies, and the lines show the mean SFRs as a function of
$M_{\rm h}$.  For the no-wind simulation in the top panels, we
reproduce the main trends of K09a: (i) SFRs at a given redshift
increase with halo mass, (ii) SFRs at a given $M_{\rm h}$ decline
with cosmic time since at least $z=4$, and (iii) cold mode star
formation dominates at all redshifts and at almost every halo mass
with the exception of intermediate masses at late times.

The lower three rows in Figure \ref{fig:sfrmode} correspond to our
outflow simulations; all have wind mode SFRs that increase steeply
with $M_{\rm h}$ (green lines), and which dominate above a certain
$M_{\rm h}$ (except for the vzw model at $z=3$).  The key thing to
note is that the wind mode begins to dominate at approximately the $M_{\rm
  h}$ where the median $t_{\rm rec}$ equals the Hubble time
(cf. Figure \ref{fig:diffrec} and $z=1$ SFRs).  Above this mass, wind
recycling provides a gas supply that quickly exceeds the
feedback-suppressed cold and hot modes.

Although outflows suppress the cold and hot modes in these
simulations, it is the mass dependence of differential recycling that
determines where and how much wind mode star formation occurs.  For
example, the cw SFR$-M_{\rm h}$ relations exhibit dramatic rises at
$M_{\rm h}\sim 10^{12} \msolar$ at all redshifts, because the median
$t_{\rm rec}$ falls below $t_{\rm H}$ above this mass.  The sw SFRs
increase steeply at $M_{\rm h}\sim 10^{11} \msolar$ for the same
reason, resulting in the global SFR density becoming wind mode
dominated at $z=2.7$ compared to $z=1.3$ for the cw (cf. Figure
\ref{fig:madau}).  In contrast, the vzw model does not show a
preferred $M_{\rm h}$ where SFR dramatically increases, because the
differential recycling curve is shallower and extends to lower masses.
For the vzw model by $z=0$, the wind mode dominates in halos above
$2\times10^{11} \msolar$, leading to more overall star formation in
sub-$M^*$ halos relative to the cw model.

The total SFRs of the wind models approach and even exceed the total
no-wind SFRs (thin black lines) at higher masses at $z=1$ and 0.  In
all our implementations, star formation-driven feedback appears unable
to prevent star formation in massive halos.  The sw SFRs are similar
to the nw SFRs above $M_{\rm h}=10^{12} \msolar$, at all redshifts.
The sw winds, launched at $340 \kms$, produce halo fountains with
$t_{\rm rec} \ll t_{\rm H}$, which result in similar star formation
rates as in the no-wind case.  In some sense, this is similar to the
canonical scenario for winds envisioned in e.g. \citet{dek86}, where
star formation-driven feedback has a strong impact on low mass
galaxies but little impact at high masses, albeit for different
reasons.

All the wind models show greater SFRs relative to the nw case in a
range of halos by $z=0$, with the sw and vzw SFRs exceeding the nw
case for $M_{\rm h}\ga 10^{12} \msolar$.  In part, the enhanced
late-time star formation reflects a delay, with wind ejection and
recycling shifting star formation from early times to late times.  In
addition, wind recycling can shift star formation from low mass
galaxies to high mass galaxies via intergalactic fountains.  Either
way, wind recycling provides a greater reservoir of gas for late-time
star formation.  The resulting wind mode-dominated SFRs in high mass
halos far exceed observed SFRs in the local Universe.

\begin{figure}
\includegraphics[scale=0.8]{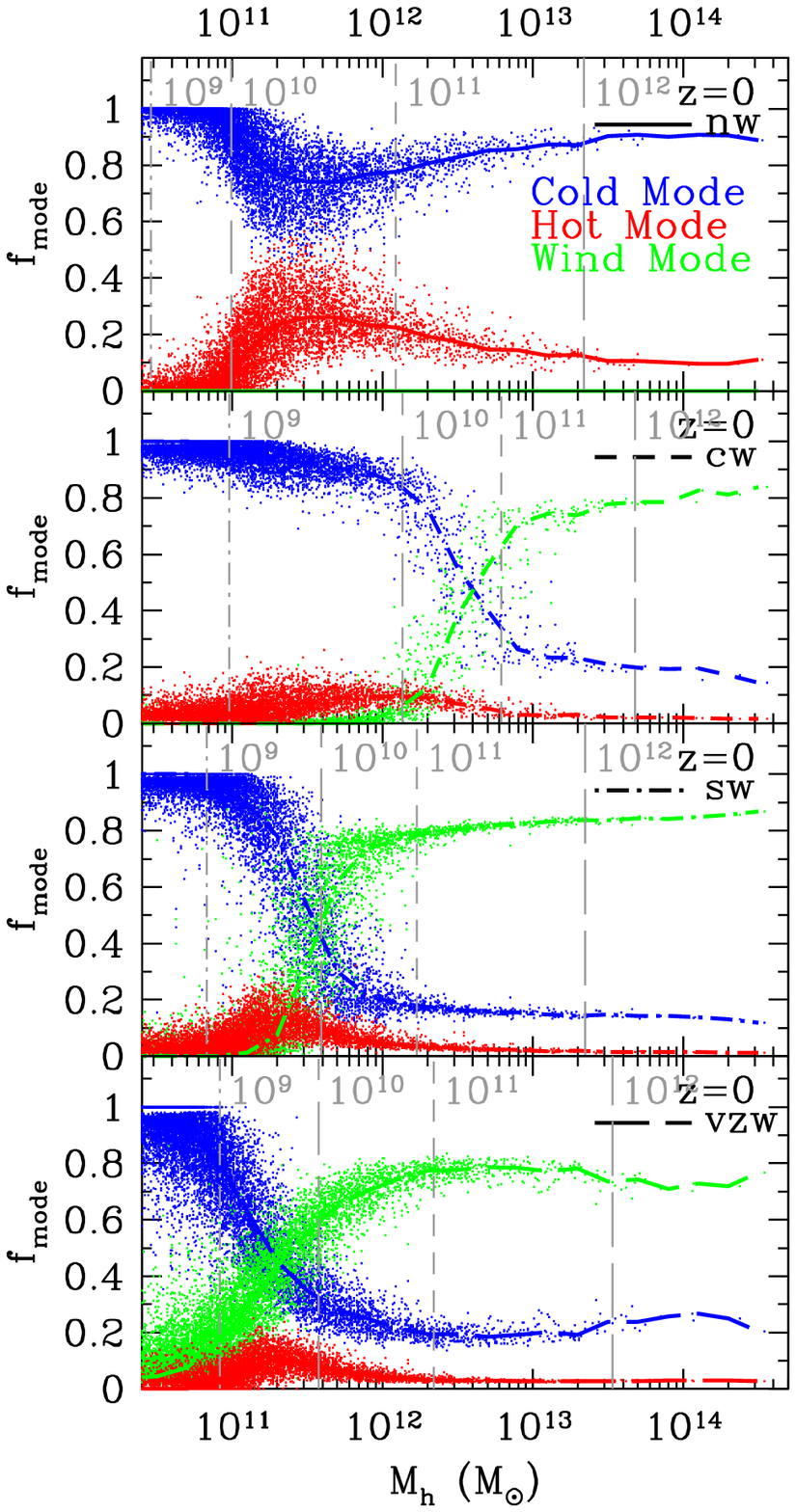}
\caption[]{The fractional stellar mass of central galaxies assembled
via the different modes as a function of halo mass.  Coloured lines
are medians of the fractional stellar mass growth corresponding to
each mode.  Cold mode dominates the no-wind simulation (top panel) at
all halo masses, always exceeding 75\% of the total mass growth.  Cold
mode exceeds hot mode in every feedback model (bottom three panels),
but wind mode becomes the dominant growth mode at high $M_{\rm h}$ in
every case.  We indicate the relation between halo and stellar mass by
grey vertical lines indicating the median $M_{\rm h}$ for the given
$M_{\rm s}$ listed next to each grey line.  }
\label{fig:fmode}
\end{figure}

Figure \ref{fig:fmode} shows the cumulative contributions of cold,
hot, and wind mode accretion to the $z=0$ stellar mass of central
galaxies.  For the nw simulation, our results are similar to K09a/b:
cold mode dominates at every mass, with a median contribution near
100\% at low masses, 75\% at $M_{\rm h} \sim 3\times 10^{11}
\msolar$, and 80-90\% above $M_{\rm h} \sim 5\times 10^{12}
\msolar$.  While direct cold mode accretion dominates at low masses,
the highest mass galaxies are built largely from mergers of low-mass
galaxies that formed {\it their} stars from cold mode gas (K09a).
To reproduce the observed (much lower) masses of these galaxies, their
low-mass progenitors must efficiently suppress star formation at
early times, likely using an ejective feedback mechanism (K09b).  

In contrast to the nw case, in all three wind models it is the wind
mode that dominates the growth of high mass galaxies.  The onset of
wind mode dominance is swift in the cw and sw models, and slower in
the vzw model owing to its shallower differential recycling curve.
The transition mass where wind mode accounts for over half of the
stellar mass growth occurs at $M_{\rm h}=10^{12.7}$, $10^{11.6}$, and
$10^{11.4} \msolar$ for the cw, sw, and vzw simulations respectively.
The corresponding transition masses for the $z=0$ SFR (right panels of
Figure~\ref{fig:sfrmode}) are lower because recycled wind star
formation occurs predominantly at late times.  We show the link
between $M_{\rm h}$ and $M_{\rm s}$ for central galaxies in Figure
\ref{fig:fmode} using vertical grey lines drawn at the median $M_{\rm
  h}$ for central galaxies of $M_{\rm s}=10^{9}$, $10^{10}$,
$10^{11}$, and $10^{12} \msolar$.  The grey lines move to higher halo
masses for the $10^{9}$ and $10^{10} \msolar$ lines with the addition
of feedback, because a galaxy with the same stellar mass will reside
in a more massive halo owing to feedback suppression.  The lines shift
much less for $10^{12} \msolar$ central galaxies as wind recycling
renders feedback ineffective at limiting stellar mass growth for
high-mass galaxies.  In terms of stellar mass, wind mode accounts for
over half of the stellar mass growth above $M_{\rm s} = 10^{10.8}$,
$10^{10.1}$, and $10^{9.8} \msolar$ for the cw, sw, and vzw
simulations respectively.

\subsection{Galactic Stellar Mass Functions} \label{sec:gsmf}

\begin{figure*}
\includegraphics[scale=0.8]{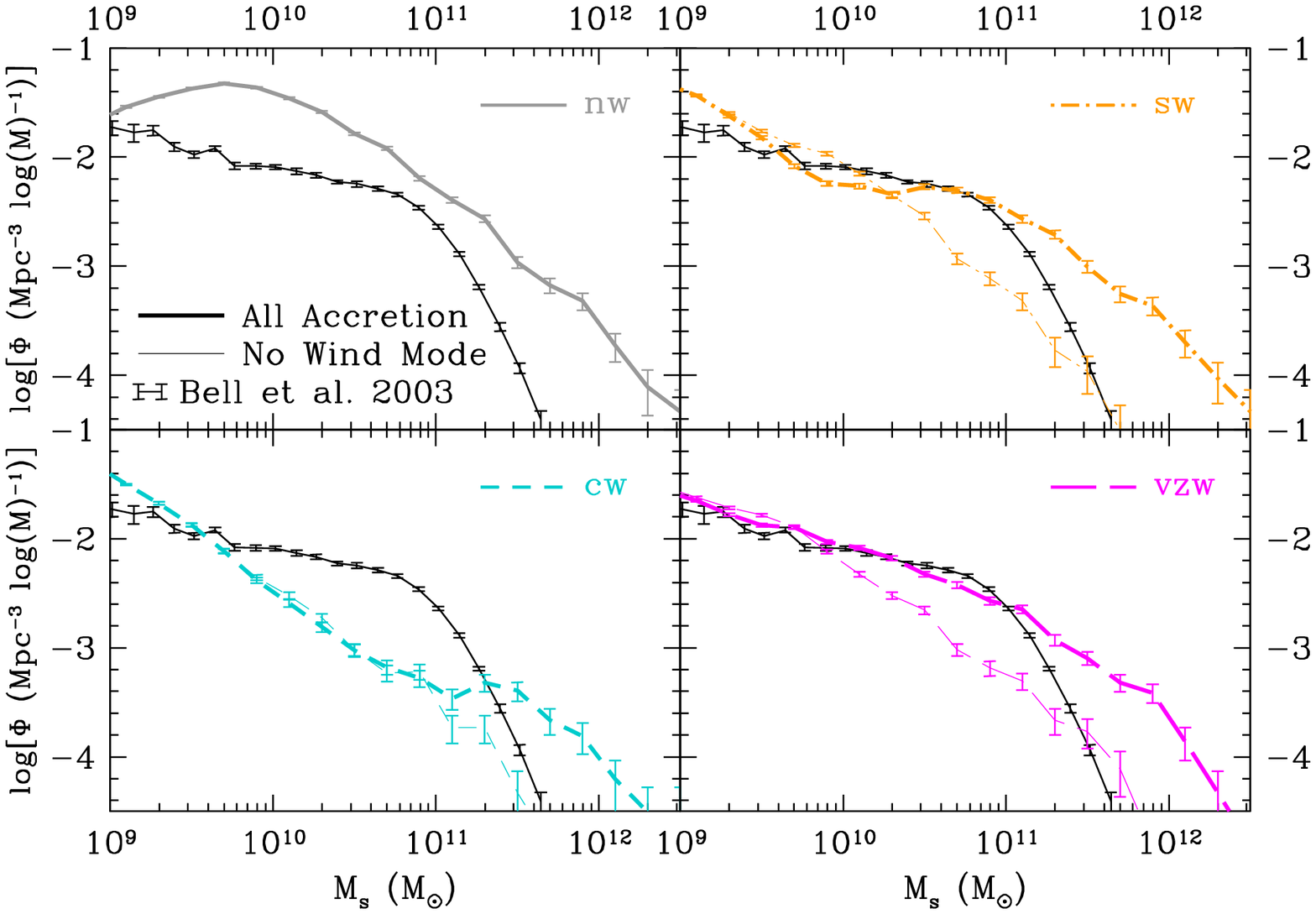}
\caption[]{The $z=0$ galactic stellar mass functions (GSMFs) for our
  four simulations including all accretion (thick lines) and without
  wind mode (thin lines); the no wind (upper left panel) simulation,
  by definition, has no wind mode.  Also shown is the \citet{bel03}
  $g$-band derived GSMF (black) scaled to a Chabrier IMF for
  comparison.  }
\label{fig:gsmf}
\end{figure*}

We now consider stellar masses for {\it all} galaxies.  Figure
\ref{fig:gsmf} shows the simulated $z=0$ GSMFs including all accretion
(thick lines) and excluding wind mode (thin lines).  The latter case
is equivalent to assuming wind materials never recycle and remain
outside the ISM of any galaxy.  Because SPH particles are tagged when
first launched, all subsequent star formation can be removed from our
analysis if particles recycle more than once.  For comparison, we show
the observed \citet{bel03} GSMF\footnote{For alternative recent
determinations of the GSMF, see \citet{li09} and \citet{ber09}.}.  As
in K09b, the nw simulation significantly over-predicts the observed
GSMF at all masses.  The cw model changes from a nearly flat power law
without wind mode ($\alpha=-2.11$ between $M_{\rm s} = 10^{9.5} -
10^{10.5} \msolar$) to having a bump above $M_{\rm s} = 10^{11}
\msolar$ including all accretion.  The result is a poor fit
everywhere.  Adding wind mode in the sw model flattens the GSMF the
most between $M_{\rm s}=10^{9.9}-10^{10.9} \msolar$, from
$\alpha=-2.18$ to $-1.11$.  The more gradual increase of wind mode
with mass changes the vzw faint-end slope from $\alpha=-1.92$ to
$-1.45$ over $M_{\rm s} = 10^{9.5}-10^{10.5} \msolar$, and provides
the best match to the observations at $M_{\rm s} \le 5\times 10^{10}
\msolar$.  Excluding wind mode yields acceptable agreement at the
highest masses for the sw and vzw models, but it spoils the agreement
at lower masses.  We discuss the relative shapes of the various GSMFs
further in \S\ref{sec:faintend}.

\begin{figure}
\includegraphics[scale=0.9]{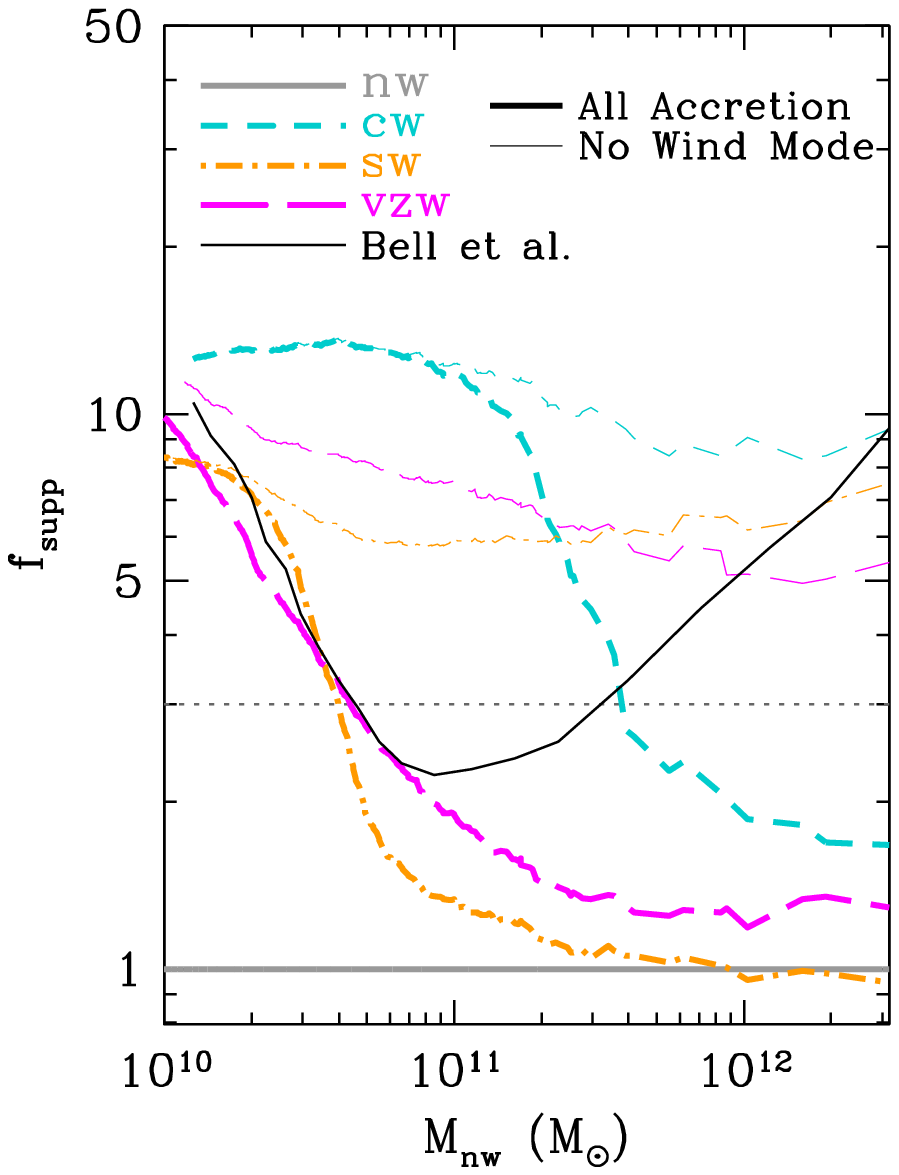}
\caption[]{The suppression factor $f_{\rm supp}$ of galaxy stellar
  masses in feedback simulations relative to the stellar masses in the
  no-wind simulation.  Galaxies are matched with the nw simulation by
  ranking them in order of decreasing stellar mass, then selecting
  galaxies of the same rank.  Thick curves show results for the cw
  (cyan), sw (orange), and vzw (magenta) models.  Thin curves show the
  results when recycled wind mode is explicitly removed.  The black
  solid curve shows the required $f_{\rm supp}$ of no-wind stellar
  masses, $M_{\rm nw}$, to match the masses of galaxies with the same
  comoving space density according from \cite{bel03}.  A model with
  $f_{\rm supp}$ following the black curve would reproduce the
  \cite{bel03} GSMF exactly.  The dotted grey line shows the
  suppression naively expected from ejective feedback in the cw and sw
  wind models when recycling is suppressed.  }
\label{fig:rank}
\end{figure}

To better characterise the impact of winds and wind recycling on
galaxy masses, we rank order the galaxies in each simulation by
stellar mass, then compare the properties of galaxies of the same rank
in the different simulations.  In a statistical sense, this procedure
picks out the directly comparable galaxies between pairs of
simulations.  Figure~\ref{fig:rank} plots the suppression factor,
$f_{\rm supp}=M_{\rm nw}/M_{\rm wind}$ relative to the stellar mass in
the no-wind simulation, $M_{\rm nw}$.  A value $f_{\rm supp}=1$ (grey
line in the figure) means that the rank-ordered masses between
simulations with and without feedback are identical, i.e. feedback has
had no effect.  The solid black curve shows the value of $f_{\rm
supp}$ required to obtain the mass of observed galaxies with the same
comoving space density according to \citet{bel03}; this is the
quantity referred to as $f_{\rm corr}$ in K09b (their figure 1).  The
no-wind simulation produces galaxies that are too massive by a factor
$\sim 2.5$ at $M_{\rm nw} \sim 10^{11} \msolar$, rising to a factor of
$5-10$ at the ends of the range plotted, similar to the findings of
K09b.  The vzw model traces this curve up to $M_{\rm nw} \sim 7\times 10^{10}
\msolar$, just as it agrees with the observed GSMF up to this mass
scale in Figure~\ref{fig:gsmf}.  However, none of our models follows
the black curve over the full mass range, just as no model reproduces
the observed GSMF at all masses.

To compare the wind models to each other, we first consider the thin
lines in Figure~\ref{fig:rank}, which show $f_{\rm supp}$ when wind
recycling is explicitly removed -- i.e. any particle that was ever in
a wind is not counted when computing the galaxies' $z=0$ masses.  The
mass suppression in the cw and sw models is considerably larger than
the naively expected factor of $1+\eta=3$ (dotted grey line), for the
reasons that will be discussed in \S\ref{sec:suppress} -- the
difference partly reflects the impact of mass recycling from evolved
stars, but the dominant effect is that winds heat the surrounding gas
and suppress accretion, providing preventive feedback in addition to
ejective feedback.  The suppression in these models is relatively
independent of mass, compared to the steadily dropping $f_{\rm supp}$
in the vzw model owing to the $\eta \propto \sigma^{-1}$ scaling of
the wind mass loading.

Including recycled wind mode (as our simulations do by default)
changes the picture radically, as shown by the thick curves in
Figure~\ref{fig:rank}.  For the cw and sw models, wind recycling has
no impact on galaxies with masses below a threshold $M_{\rm nw}$.  For
the slow winds of the sw model, $f_{\rm supp}$ approaches one above
$M_{\rm nw}\approx 10^{11} \msolar$.  The cw model shows a similar
behaviour at an $\sim8 \times $ higher mass, but these faster winds
suppress the masses of even the largest galaxies by a factor of two.
The transition between the low and high mass regimes occurs at
approximately the scale where $t_{\rm rec} \approx t_H$ in
Figure~\ref{fig:diffrec}.  $f_{\rm supp}$ in the vzw model also
approaches one above $M_{\rm nw}\approx 2 \times 10^{11} \msolar$, but
it continues to rise all the way to small masses unlike the constant
wind models, which flatten towards the smallest masses.

The full $f_{\rm supp}$ curves show a much stronger mass dependence
and much larger differences from model to model than the curves
without wind mode.  We conclude that it is the mass dependence of the
recycled wind mode, what we have called differential recycling, that
plays the strongest role in determining the shape of the GSMF in our
simulations.  In all the models, wind recycling leads to excessive
stellar masses and star formation rates in massive galaxies compared
to those observed.  However, simply suppressing all wind recycling
does not lead to an agreement with the observed GSMF.  A model that
resembled vzw, including recycling, up to $M_{\rm s}\sim 10^{11}
\msolar$, but suppressed wind recycling with increasing effectiveness
at higher masses would better match observations.

\section{Discussion}

\subsection{How Feedback Suppresses Star Formation}\label{sec:suppress}

As noted in our discussion of Figure~\ref{fig:rank}, in the absence of
wind recycling, stellar mass suppression in the sw and cw models is
considerably stronger than the naively expected factor of $1+\eta=3$.
One may expect ejective feedback to reduce star formation by two
thirds when $\eta=2$, because two $M_\odot$ of mass is ejected for
every one $M_\odot$ of stars formed.  Part of the difference reflects
the impact of stellar mass recycling (distinct from wind recycling).
With our Chabrier IMF, 18\% of the mass that forms stars is promptly
recycled to the ISM as supernova and massive star ejecta, and over 10
Gyr this fraction rises to $\sim 50\%$, mostly from AGB winds.  In a
no-wind model, this recycled gas generally forms stars, increasing the
galaxy stellar mass.  However, in a constant-wind model, a fraction
$\frac{\eta}{1+\eta}$ of the initial gas supply is ejected and hence
never contributes any recycled stellar mass, and $\frac{\eta}{1+\eta}$
of the gas that is returned to the ISM by evolved stars is ejected
rather than forming stars.  With stellar recycling fractions of 50\%,
this effect alone can raise the mass suppression factor from 3 to
$\sim5$.

Furthermore, winds inject energy and momentum into their surroundings
and thereby suppress accretion, providing preventive feedback in
addition to ejective feedback.  This preventive feedback has the
greatest impact on galaxies below $M^*$.  To investigate this, we ran
a cw simulation where we did not allow winds to interact
hydrodynamically at all.  The stellar masses {\it increased} by a
factor in excess of $2$ for the smallest galaxies we resolve.  The
energy added by these strong winds to the surrounding gas via
hydrodynamic interactions is evidently playing a key role in
suppressing gas accretion.  This local ``pre-heating'' prevents IGM
material from joining the filaments through which accretion is
efficient.  However, wind pre-heating is far weaker in the more
moderate wind prescriptions.  With only a quarter as much energy
imparted to the surrounding gas, the sw winds result in galaxies twice
as massive below $M_{\rm nw}=10^{11} \msolar$ compared to the cw
prescription (cf. the thin orange and cyan lines in Figure
\ref{fig:rank}).  The vzw winds have even lower wind energies at the
low-mass end.  Nevertheless, in all wind models cold mode SFRs decline
faster at late times than expected from ejective suppression alone,
likely owing to more dilute gas that can more easily be affected by
energy input.

The vzw case also behaves somewhat differently than simple
expectations.  Outflow suppression does not rise as significantly as
one would expect toward higher masses (thin magenta line in Figure
\ref{fig:rank}), despite the $\eta\propto\sigma^{-1}$ scaling.  If we
assume $\sigma\propto M_{\rm b}^{1/3}$, then $\eta \propto M_{\rm
b}^{-1/3}$ (at a given redshift), which OD08 showed is a fair
description for the vzw mass feedback rates.  Over the range of
$M_{\rm nw}=10^{10.3}-10^{12.3} \msolar$, we would expect an increase
of $f_{\rm supp}$ by about a factor $4-5$ in the vzw suppression
efficiency relative to the constant mass loading of the cw and sw
models given how $\eta$ is calculated. However the increase appears to
be at most a factor of $2$, mainly because the more massive galaxies
are the hierarchical merger products of low-mass progenitors, which
had higher $\eta$ before merging.  This is the same argument K09a used
to explain why massive galaxies are mainly assembled via cold mode --
their progenitors are multiple small galaxies receiving gas via the
cold mode channel.  Thus, the higher $\eta$ of the progenitors balance
the low $\eta$ of the merged galaxy at late times, reducing the
differential mass loading effect.

These results show that, even in the absence of the recycled wind
mode, the suppression of stellar masses does not follow simple
scalings with the mass loading factor $\eta(M_{\rm b})$.  Stellar mass
recycling, hierarchical galaxy assembly, and, above all, the
preventive feedback effects of winds (wind pre-heating), all act to
complicate the mass dependence of feedback suppression.  These results
suggest caution when using analytic or semi-analytic models tied to
the observed GSMF to infer the scaling of $\eta$ with galaxy mass.
Furthermore, we find that wind recycling changes the mass dependence
of feedback suppression more strongly still, so model predictions will
be sensitive to the assumptions (explicit or implicit) about the fate
of gas after it has been ejected from the star-forming regions of its
host galaxy.

\subsection{Recycling and Faint-End Slopes of the GSMF} \label{sec:faintend}

Differential recycling and wind mode creates much more diversity
between the slopes of the mass functions (cf. thick coloured lines in
Figure~\ref{fig:gsmf}) than feedback suppression alone (thin lines).
The sw model results in three distinct mass regimes, each with a
different slope.  At $M_{\rm s}\la 10^{10} \msolar$ the slope remains
steep until the wind mode significantly softens the slope as it
increases sharply between $M_{\rm s}\sim 10^{10}-10^{11} \msolar$.
Above $M_{\rm s}\ga 10^{11} \msolar$ the slope steepens again as
efficient recycling causes it to approach the nw slope.  This triple
slope behaviour has also been identified in the OverWhelmingly Large
Simulations (OWLS) when using a constant wind-type model (Marcel Haas,
private communication, see also \citet{cra09}).  The cw mass function
could be described in a similar way, although with higher $\vw$ the
``flattening'' from the wind mode is better described as a bump at
high masses.  Increasing $\vw$ by a factor of two dramatically alters
the $z=0$ GSMF in two main ways: (i) preventive feedback suppresses
by $\sim \times 2$ cold and hot mode stellar growth at most masses,
and (ii) wind mode galaxy growth begins at $\sim \times 8$ higher
mass.

The vzw simulation manages to follow the faint-end slope of
\citet{bel03} between $M_{\rm s}=10^{9}- 10^{10.7} \msolar$ fairly
well.  The flatter faint end is a direct result of the gradual
differential recycled wind mode inherent in the vzw model.  Above the
threshold mass where recycling dominates, outflows behave as a series
of fountains eventually becoming unable to escape their parent halo
(i.e. the halo fountain phenomenon at $\ga M^*$, OD08).  Below the
threshold mass where winds mainly escape into the IGM, wind mode
contributes less to galaxies resulting in a very subtle steepening of
the GSMF below $10^{9.5} \msolar$, a milder version of the
``triple-slope'' behaviour in the sw model.  Our resolution limit
prevents us from concluding whether this simulated GSMF reproduces the
``faint-end upturn'' relative to a Schechter function, hinted at by
the \cite{bel03} data and seen more clearly by \cite{bal08}; however
we suggest the possibility this upturn could represent the transition
mass below which wind mode stops contributing and the slope steepens.

\begin{figure}
\includegraphics[scale=0.8]{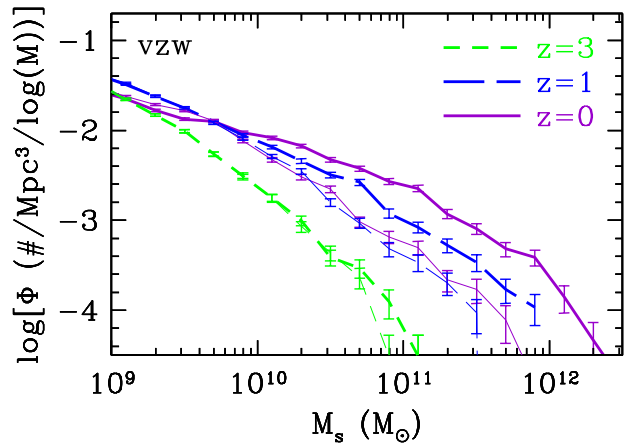}
\caption[]{The GSMFs of the vzw model from $z=3\rightarrow 0$
including all accretion (thick lines) and without wind mode (thin
lines).  }
\label{fig:mstar_evol}
\end{figure}

Differential recycling also results in an evolution in the faint end
of the GSMF, such that it is steeper at high redshifts when recycling
is a small effect compared to direct accretion.  Figure
\ref{fig:mstar_evol} demonstrates this evolution for the vzw model.
Feedback sets the slope of the
GSMF at all epochs; the steeper slopes at high-$z$ ($z\ga 2$) are set
by the ejection of gas while the flatter slopes at late times are set
by the recycling of gas.  This evolution in the faint end has long
been a challenge for hierarchical models, with only relatively ad hoc
explanations proposed~\citep[see e.g.][]{kho07}.  It is encouraging
that our vzw wind model can fit the observed high-$z$ GSMFs in
\citet[][see also \citealt{dav06}]{opp09b}, while concurrently showing
a number of other successes over other wind models in reproducing key
galaxy and IGM statistics (see \S1).

\subsection{Is Wind Recycling a Numerical Artifact?} \label{sec:numeff}

The detailed hydrodynamic and radiative interactions that might drive
a galaxy's ISM into outflows occur on scales well below the resolution
limit of any SPH simulations that model a representative cosmological
volume.  As discussed in \S\ref{sec:sims}, \gad\ uses a ``sub-grid''
model to store thermal energy from stellar feedback in the hot phase
of a two-phase ISM, which slows star formation, pressurises and
stabilises the ISM, and improves numerical convergence by lowering the
sensitivity to mass resolution \citep{spr03a}.  However, this ISM
pressure does not, on its own, lead to galactic-scale outflows, so
\gad, like some of its predecessors \citep[e.g.][]{nav96}, implements
winds by imparting kinetic energy to individual particles
\citep{spr03a}.  Winds are temporarily hydrodynamically decoupled as
described in \S\ref{sec:sims} to simulate the formation of hot
chimneys extending out of a disk galaxy and to obtain resolution
convergence in feedback simulations \citep{spr03b}.  Without this
decoupling, it is still possible to drive winds that suppress star
formation \citep{sch10} and enrich the IGM \citep{wie09a}, but the
results do not remain well-converged with resolution
\citep[e.g.][]{dal08}.  We have modified the \gad~algorithm to
implement the velocity scaling of momentum-driven winds (following
OD08), but we have not altered the basic mechanism by which wind
feedback is implemented.  Our exploration of feedback in this paper is
limited to kinetic feedback only, and other methods of feedback
\citep[e.g. SN blastwaves, ][]{sti06} will result in different answers.

By construction, our scheme achieves its original goal, ejecting gas
from star-forming galaxies with mass-loading factors and wind
velocities that are motivated by a combination of theoretical and
empirical arguments.  Our results (and those of OD08) show that {\it
re-accretion} of these ejected wind particles plays a major role in
shaping the distribution of galaxy stellar masses and the correlations
among halo mass, stellar mass, SFR, and redshift.

Unfortunately, there are a variety of numerical effects that could
cause our modelling of the post-ejection behaviour of wind particles
to be inaccurate.  Most obviously, ejected wind particles initially
have velocities and temperatures that are very different from those of
their neighbours in the galaxy halo, while SPH is designed to represent
fluid elements by groups of $N\geq 32$ particles.  Second, SPH has
known difficulties in treating two-phase media with sharp
discontinuities in temperature and density, where gaps that develop
around dense clumps tend to suppress instabilities that might
otherwise break them apart \citep{age07}.  Third, the ram pressure on
under-resolved clumps (or single particles) can be overestimated
because their smoothing lengths must be large enough to enclose 32
neighbours, which may give them an artificially large cross-section.
K09a suggest that this excess ram pressure causes some or all of the
``cold drizzle'' of $T \approx 10^4$ K gas onto the central galaxies
of massive halos \citep[see also][]{naa07}.  Fourth, our calculation
of metal-line cooling is based on collisional equilibrium tables while
\cite{wie09a} showed that photoionisation can significantly increase
metal-line cooling times.  In simulations without metal-line cooling,
we find a significant decrease in wind recycling, and in overall gas
accretion for galaxies above $M^*$, in general agreement with the
findings of \cite{cho09}.  Conversely, including the mixing of metals
via instabilities, conduction, or diffusion \citep[e.g.][]{wad08}
would generally increase the overall accretion in hot halos, as metal
coolants spread over more mass decrease cooling times.  \citet{wie09b}
demonstrated that using smoothed SPH metallicities to simulate metal
mixing can increase by 50\% the number of baryons in stars compared to
using single particle metallicities.

Wind recycling is unquestionably an important effect in our
simulations, as we have shown in the preceding sections.  It is a
physically plausible phenomenon that doubtless occurs to some degree
in the real Universe.  However, given the numerical issues listed
above, we cannot be sure that our simulations give an accurate
quantitative account of the level of wind recycling that should arise
in our adopted physical scenarios (cw, sw, and vzw).  These numerical
issues should be less severe for lower mass galaxies without hot
halos, therefore the predictions for recycling relating to the
sub-$M^*$ galaxies and the GSMF faint-end slope should be more robust.
It is worth reiterating that simulations with the \gad~wind algorithm
and the vzw scalings have achieved a number of empirical successes,
including the $z=6$ galaxy luminosity function \citep{dav06}, the
galaxy mass-metallicity relation \citep{fin08a}, observations of IGM
metal enrichment over a wide range of redshifts
\citep{opp06,opp09a,opp09b}, and the shape of the $z=0$ GSMF below
$M^*$ (this paper).  These agreements do not depend much on the
recycling behaviour in high-mass galaxies.  Conversely, wind recycling
contributes to the excessive masses and SFRs of super-$M^*$ galaxies
in high-mass halos, which is partially why the same simulations
over-predict the global SFR density below $z\la 1$ (Figure
\ref{fig:madau}) and fail to reproduce the locally observed colour
bimodality of galaxies \citep{gab10}.

In comparing simulations with different mass resolutions ($2\times
256^3$ particles in 16, 32, and $64\hmpc$ boxes), OD08 found similar
qualitative results for wind recycling but significant changes in
$t_{\rm rec}$ at a fixed galaxy mass.  Further investigations of
resolution and box size effects will be needed to characterise the
numerical robustness of our quantitative results.  In a different
approach, we are using very high resolution simulations with idealised
geometries to compare the behaviour of single-particle ejection to
ejection of well resolved gas blobs (M.\ Peeples et al., in
preparation).  The strongest test of numerical robustness would be to
carry out simulations that implement momentum-driven winds with a
substantially different numerical algorithm; this is a longer term
goal.  At the same time, one can search for more targeted
observational tests of our numerical predictions for galactic outflows
and wind recycling, a point we return to below.

\section{Conclusions}

We have examined the factors that govern the present-day galaxy
stellar mass function in cosmological hydrodynamic simulations that
incorporate several different galactic outflow prescriptions.  Our key
result is the identification of a mode of accretion that we call
recycled wind mode (or just wind mode), in which the accretion comes
from material that was previously ejected from a galaxy.  This
provides a third distinct accretion mode along with the ``cold'' and
``hot'' modes described by \citet[e.g.][]{ker05,dek06}.  At some
level, the existence of a recycled wind mode is obvious -- material
ejected from galaxies can be subsequently re-accreted -- but the
impact on our simulations is remarkably large.  In our simulation
without winds, we recover the results of Kere\v{s} et al. (2009a/b),
in which cold mode accretion dominates the growth of galaxy masses.
In simulations with winds, however, wind recycling is the dominant
mechanism of galaxy growth at $z\leq 1$.

Wind mode accretion occurs despite winds being ejected primarily at
velocities exceeding the escape velocity of the galaxies' halos.
Hydrodynamic processes are mostly responsible for slowing the winds
(as also argued by OD08).  Even when the outflow speed is proportional
to galaxy velocity dispersion, winds recycle back into more massive
galaxies faster because they live in environments of higher gas
density.  The mass dependence of the wind mode we call {\it
differential wind recycling}.  In all three of our wind models, the
median recycling time declines at higher halo masses, but the rate of
decline depends on the wind prescription.  In the most massive halos,
the recycling time is typically below 1~Gyr, and $>90\%$ of wind
particles recycle without ever leaving their parent halos.

The differential wind recycling relations (\S\ref{sec:diffrec}) are
mirrored in the galaxy SFRs decomposed by accretion mode
(\S\ref{sec:SFRs}).  Wind mode dominates the SFR of central galaxies
approximately when $t_{\rm rec}$ becomes less than the Hubble time, so
that the outflow material is then accreted back into galaxies at a
rate exceeding the cold mode inflow.  This transition to wind mode
dominance is reflected in the stellar mass growth of $z=0$ galaxies,
as the galaxy stellar mass functions (GSMFs) from our three wind
models show significant differences that are directly traceable to the
behaviour of the wind mode in each case.

If we explicitly eliminate wind mode accretion in post-processing --
by not counting recycled particles when computing galaxy stellar
masses -- then the $z=0$ GSMFs of our three wind models are
surprisingly similar to each other, and completely unlike the observed
GSMF.  The suppression relative to the no-wind simulation is much
larger than the naively expected factor of $(1+\eta)$, more so when
high-velocity winds heat the surrounding gas and suppress subsequent
accretion, providing preventive feedback in addition to the ejective
feedback.  When we include wind recycling (as our simulations do by
default), then the GSMFs of the three wind models become quite
different from one another, and the GSMF of the momentum-driven
feedback (vzw) model agrees best with the \cite{bel03} data for galaxy
stellar masses $10^9 \msolar \leq M_{\rm s} \leq 5\times 10^{10}
\msolar$.  The addition of wind mode flattens the faint-end mass slope
from $\alpha=-1.92$ to $-1.45$ between $10^{9.5} \msolar \leq M_{\rm
s} \leq 10^{10.5} \msolar$.  However, the predicted GSMFs greatly
exceed the observed GSMF for $M_{\rm s}\geq 10^{11} \msolar$,
approaching the prediction of the no-wind simulation, because material
ejected in winds is quickly re-accreted in high mass halos.

Reproducing the observed $z=0$ GSMF through wind mode alterations
alone, at least in the vzw model, would require strong suppression of
wind recycling at super-$M^*$ masses ($M_{\rm h} \ga 2\times 10^{12}
\msolar$), but minimal suppression of wind recycling at lower masses.
It is conceivable that AGN feedback could provide a mechanism for such
mass dependent suppression, just as it has been invoked to suppress
hot mode accretion in high mass halos (e.g., \citealt{cro06, cat09}).
However other mechanisms including virial shock heating \citep{bir07,
naa07} and gravitational heating by infalling satellites and cold gas
clumps \citep{dek08, kho08} may suppress star formation without AGN,
although the resolution in our cosmological-scale simulations may be
insufficient to resolve such processes.  Using semi-empirical modeling
\citet{hop08} demonstrated that if star formation is efficiently
quenched after major mergers of gas rich disks then the fraction of
passive galaxies as a function of stellar mass is well matched to
observations.  \citet{gab10} showed quenching mechanisms applied
post-run to vzw simulations can produce a red sequence of galaxies
without dramatically changing galaxies below $L^*$, although in detail
none of the mechanisms tested there can exactly reproduce both
observed luminosity functions and color-magnitude diagrams.  Even if
all star formation is quenched at $M^*$ and above, this cannot account
for the factor of $\sim\times 5$ greater SFR density than observed.
Furthermore, \citet{fir09} demonstrated recycling preferentially
increases the specific SFR (SFR/$M_{\rm s}$) in massive galaxies,
which is in contradiction with observations showing the opposite
trend.  Simulations including self-consistent mechanisms to quench
massive galaxy star formation may yield unexpected consequences for
the star-forming sequence of galaxies.

We note that the impact of wind mode is particularly strong at late
cosmic epochs, while at earlier times ($z\ga 2$) the GSMF faint-end
slope is set more by ejection.  In our favoured wind model this causes
the faint end to become shallower with time, which is qualitatively in
agreement with observations.  We leave a more detailed examination of
such evolutionary trends for future work.

While wind recycling is an important effect in our simulations, we
discussed (in \S\ref{sec:numeff}) several reasons that our numerical
treatment of winds and wind recycling could be inaccurate.  Assessment
of our predictions will require further testing of our sensitivity to
numerical resolution, a study of the performance of the \gad\ wind
implementation (particularly the single-particle ejection scheme), and
consideration of alternative treatments of metal-line cooling.  We can
also look for direct observational tests of our numerical predictions
of wind recycling and the distribution of ejected gas in galactic
halos.  One indirect line of evidence comes from the galaxy
mass-metallicity relation \citep[e.g.][]{tre04}, which implies a steep
dependence of mass-loading factor on halo virial velocity, roughly
$\eta \propto v_{\rm vir}^{-3}$ (M. Peeples \& F. Shankar, in
preparation; R. Wechsler, private communication).  This dependence,
steeper than that expected for momentum- or energy-conserving winds,
could potentially be explained by a combination of the $\sigma
^{-1}$ scaling for momentum-driven winds with the mass-dependent
recycling rates that cause more massive galaxies to re-accrete much
more of the enriched gas they once ejected.  More direct tests of the
simulation's predictions can come from quasar absorption studies that
probe metals and ionization states in the environments around
galaxies.  The advent of the {\it Cosmic Origins Spectrograph} on {\it
Hubble Space Telescope} allows much more sensitive searches for
low-redshift metals, including tests of the \citet{opp09a} prediction
that most metals traced by $\OVI$ reside near galaxies and frequently
recycle.

Modellers have long argued that galactic outflows are a crucial
element of galaxy formation, necessary to understand the shape of the
galaxy luminosity function and the overall low efficiency with which
cosmic baryons are converted into stars
\citep[e.g.][]{whi91,kau93,col94,som99}).  Hydrodynamic simulations
that explicitly incorporate winds into them are more successful than
those without winds at reproducing other aspects of galaxy evolution
and IGM enrichment.  Those with moderate outflows from the majority of
galaxies both effectively suppress star formation and enrich the
nearby IGM where most metals are observed to reside
\citep[e.g.][]{sto06, wak09}.  Our results show that these outflows
are by no means the end of the story -- to understand how galaxies
grow, one must understand, theoretically and observationally, the fate
of the gas that they expel.

\section*{Acknowledgments}  \label{sec: ack}

We thank the referee, Avishai Dekel, for numerous suggestions that
improved and clarified the paper.  The authors also wish to thank Ari
Maller, Kristian Finlator, Jared Gabor, Molly Peeples, Joop Schaye,
Craig Booth, Marcel Haas, Ryan Quadri, Rob Wiersma, George Becker, and
Richard Bower for encouraging discussion related to this research.
The simulations run on the Intel 64 Linux Cluster Abe Supercluster at
the National Center for Supercomputing Applications were greatly aided
by the support staff there by providing us with dedicated nodes.  We
also thank the support staff for the ICE SGI cluster at the University
of Arizona and Craig West at the Eagle Opteron cluster at the
University of Massachusetts.  Support for this work was provided by
NASA through grant number HST-AR-10946 from the Space Telescope
Science Institute.  DW acknowledges the hospitality of the Institute
for Advanced Study and the support of an AMIAS Membership.

\label{lastpage}

\end{document}